\documentclass[10pt,aps,prapplied,twocolumn,superscriptaddress, longbibliography, amsmath,amssymb]{revtex4-2}
\usepackage{graphicx}
\usepackage{color}
\usepackage{ulem}
\usepackage[bookmarks,colorlinks,breaklinks]{hyperref} 
\hypersetup{linkcolor=red,citecolor=blue,filecolor=dullmagenta,urlcolor=blue} 
\hypersetup{pdftoolbar=true,pdfpagemode=UseNone,pdfstartview=FitH, colorlinks=true,extension=}
\renewcommand{\Re}{\textrm{Re}}
\renewcommand{\Im}{\textrm{Im}}
\begin{document}
\title[BOUND STATES IN THE CONTINUUM INDUCED...]{Bound states in the continuum induced via local symmetries in complex structures}

\author{Cheng-Zhen Wang}
\affiliation{Wave Transport in Complex Systems Lab, Department of Physics, Wesleyan University, Middletown, Connecticut 06459, USA}
\author{Ulrich Kuhl}
\affiliation{Université Côte d’Azur, CNRS, Institut de Physique de Nice (INPHYNI), 06200, Nice, France}
\author{Adin Dowling}
\affiliation{Wave Transport in Complex Systems Lab, Department of Physics, Wesleyan University, Middletown, Connecticut 06459, USA}
\author{Holger Schanz}
\email[]{holger.schanz@h2.de}
\affiliation{University of Applied Sciences Magdeburg–Stendal, 39114 Magdeburg, Germany}
\author{Tsampikos Kottos}
\email[]{tkottos@wesleyan.edu}
\affiliation{Wave Transport in Complex Systems Lab, Department of Physics, Wesleyan University, Middletown, Connecticut 06459, USA}

\begin{abstract}
Bound states in the continuum (BICs) defy the conventional wisdom that assumes a spectral separation between propagating waves, that carry energy away, and spatially localized waves corresponding to discrete frequencies.
They can be described as resonance states with infinite lifetime, i.e., leaky modes with zero leakage.
The advent of metamaterials and nanophotonics allowed the creation of BICs in a variety of systems.
Mainly, BICs have been realized by destructive interference between outgoing resonant modes or by exploiting engineered global symmetries that enforce the decoupling of a symmetry-incompatible bound mode from the surrounding radiation modes.
Here, we study BICs relying on a different mechanism, namely {\it local} symmetries that enforce a field concentration on a part of a complex system without implying any global symmetry.
We experimentally implement these BICs using microwaves in a {\it compact} one-dimensional photonic network.
We demonstrate that such BICs form an infinite ladder in k-space and emerge from the annihilation of two topological singularities, a zero and a pole, of the measured scattering matrix.
This alternative for achieving BICs in complex wave systems may be useful for applications such as sensing, lasing, and enhancement of nonlinear interactions that require high-$Q$ modes.
\end{abstract}

\maketitle
\section{Introduction}\vspace{-0.5cm}

Quantum mechanics books, typically distinguish between two type of states: bound states whose discrete energy lies below the continuum threshold (identified by the asymptotic value of the potential at infinity) and unbounded scattering states with an energy inside the continuum.
Examples where these two categories of states appear include electrons in the presence of finite potential wells, quantum dots and or atomic potentials.
An exception to the quantum {\it communis intellectus} are bound states in the continuum (BICs) \cite{HZSJS16,S21,JPSJ21,AK21,M09,S69,SH75}.
These are spatially bounded solutions of the Schr\"odinger equation, with discrete eigenvalues lying inside the continuum of states that propagate to infinity.
They were originally introduced a century ago by von Neumann and Wigner, using an inverse potential engineering approach \cite{NW29}.
The method assumed a square-integrable BIC wave function with a spatially decaying envelope and, using this as a starting point, they tailored a suitable three dimensional (3D) potential where this wave function is an eigenmode.
Such ``custom-made'' potentials are unrealistic as they are oscillatory in space while decaying to infinity according to a power law.
They have never been realized.

Actual experimental implementations of BICs so far have relied mostly on extended systems.
In fact, there is a nonexistence theorem for compact structures \cite{HZSJS16} (see, however, Refs~\cite{Law+21,D+22} for exceptions).
Finding genuine BICs in compact systems therefore is a challenging fundamental problem, and its solution would allow for high-$Q$ resonators with a broad range of potential applications including lasers, sensors, filters, and low-loss fibers \cite{HZSJS16,S21}.

BICs are not exclusively a quantum phenomenon but rather pertain to all wave systems.
This observation extended the search for BICs to a variety of other platforms including electromagnetic, acoustic, water, and elastic waves (for a review see \cite{HZSJS16,S21}).
Among the various areas, optics and photonics have undoubtedly been the tip of the spear as far as novel realizations of BICs are concerned \cite{ZHLSS14,S21,GXF17,HZSJS16,JPSJ21,AK21,MBS08,PPDHNSS11,WSZ22,DMHAK18}.
For example, an inverse design scheme based on supersymmetric transformations has been implemented in coupled optical waveguide arrays to engineer the hopping rate between nearest resonators in order to support BICs \cite{CVCOL13}.

\begin{figure*}
\centering
\includegraphics[width=\linewidth]{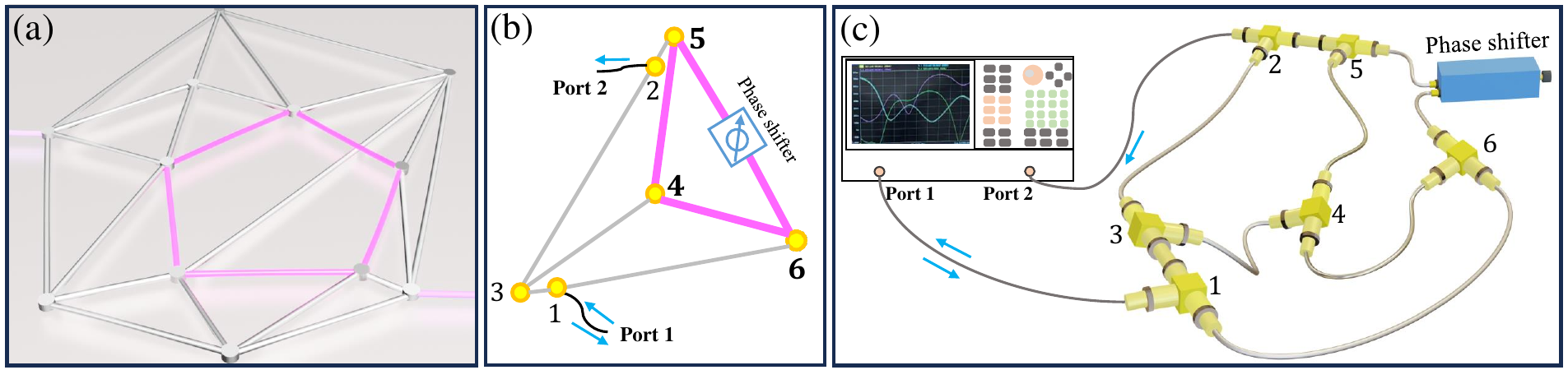}
\caption{ \label{fig:ExpSetup}
	{\bf Experimental setup of a microwave graph detecting bound states in the continuum (BICs) based on local symmetries.
	}
    (a) Schematic representation of a complex network supporting a BIC due to the presence of a subnetwork (cycle) with a local symmetry (red-shaded equilateral pentagon).
    (b) Schematic representation of the network used in our experiment.
    A "tetrahedron" contains an equilateral triangle where one side length can be varied.
    The network is opened by two attached scattering channels.
    (c) The microwave tetrahedron network used in our experiments.
    Coaxial cables are connected by T-junctions at each of the vertices $n=1, 2, \dots, 6$.
    Between vertices $5$ and $6$, we replace the cable with a phase shifter allowing us to vary the effective length between these vertices.
    The reflection amplitudes $r_{11}$, $r_{22}$ and the transmission amplitudes $t_{12}$, $t_{21}$ are measured using a vector network analyzer (VNA) connected to vertices 1 and 2, respectively.
}
\end{figure*}

Other, more efficient, BIC schemes have been also developed and experimentally implemented thanks to the advent of metamaterials and nanotechnology \cite{HZSJS16,S21,JPSJ21,AK21}.
Perhaps the most prominent approach, which also highlights the wave nature of the phenomenon, is associated with ``accidental'' BICs \cite{FW85a,FW85}.
In this case, the parameters of the target system are fine-tuned to achieve cancelation of outgoing waves to the continuum.
A special case of such accidental BICs is associated with Fabry-Perot configurations \cite{MBS08,HZCJJS13,WXKMTNSSK13} of two identical resonances which interact strongly through the same radiation channel (e.g.\ a waveguide).
A third mechanism that leads to BICs invokes structures with geometric symmetries \cite{Rob86,HZSJS16,S21}.
In this case, a trapped mode with a given symmetry can be embedded into a continuum of states with a distinct symmetry, ensuring the decoupling of the trapped mode and thus the suppression of its leakage.

A global symmetry is a strong restriction excluding the vast majority of systems, in particular when complex environments are involved.
Therefore it is crucial to realize BICs with the weaker requirement of a local symmetry.
``Local symmetry'' means that a symmetric substructure is embedded in an asymmetric environment; see for example Fig.~\ref{fig:ExpSetup}(a).
Such situations have recently received a lot of attention \cite{S+20,RMS18,K+13b,K+14,R+19a}.
In particular, it has been shown that local symmetries can be realized in photonic systems \cite{S+20} and that they can be used to create localized states in extended periodic and aperiodic systems \cite{Sgrignuoli19} and in scattering systems, i.e.~BICs \cite{RMS18}.
The latter, however, has been restricted to theoretical studies.
Here we realize for the first time BICs based on local symmetries experimentally.
We do so using a complex network of coaxial microwave cables coupled together via T-junctions, see Figs.~\ref{fig:ExpSetup}(b) and \ref{fig:ExpSetup}(c).
Networks of wave guides are relevant in various contexts and experimental implementations have been reported in a variety of frameworks including acoustics \cite{RRTMSHA23}, microwaves \cite{H+04c,RAJMSKS16,Chen22,L+20,All+2014,CKA20}, photonic crystal waveguides \cite{SAA22}, and optics \cite{NPCK23,Lepri17}.

We identify the signature of BICs in the scattering data for a two-terminal setup.
An interesting feature is the appearance of an infinite ladder of BIC states which occur periodically in k-space.
The analysis of the experimental scattering matrix demonstrates that the formation of BICs originates from the coalescence of two topological defects with opposite charges: a zero (with charge $+1$) and a pole (with charge $-1$) of the scattering matrix.

\section{Physical mechanism for local-symmetry protection of BICs}
Let us explain the {\it local-symmetry} protection mechanism using complex photonic networks as an example [Fig.~\ref{fig:ExpSetup}(a)]: the subdomain (subnet) that possesses a local symmetry is formed by a closed loop (ring) of $N_{l}$ equilateral edges such that mirror symmetry (with respect to the vertices defining the subnet) and discrete rotation invariance are guaranteed (see, for example, the violet pentagon in Fig.~\ref{fig:ExpSetup}(a) with $N_l=5$ and the violet triangle in Fig.~\ref{fig:ExpSetup}(b) with $N_l=3$ having a discrete rotational symmetry $C_{5v}$ and $C_{3v}$, respectively).
Note that in the case of a network that is formed by photonic waveguides, the geometric shape of the edges and the angles between them are typically irrelevant; see Fig.~\ref{fig:ExpSetup}(c).
In this case the discrete rotation must not be interpreted in physical space.
For example, in case of Fig.~\ref{fig:ExpSetup}(b), an angle is defined as $2\pi x/(3\ell)$, where $x$ is a continuous path length along the cycle, $\ell$ is the length of one cable and $x=0$ coincides with a vertex.
Then the subnet that supports a BIC is invariant under the transformations $x\to x+\ell$ and $x\to -x$, generating the symmetry group $C_{3v}$.

Consider the ring first without any connections to the remaining network.
Then, choosing the appropriate symmetry class, there are rotation-invariant eigenfunctions which are antisymmetric with respect to the vertices.
That is, we have eigenfunctions vanishing at all vertices of the subnet.
Thus, when the remaining network is coupled to some or all vertices of the subnet via coupling constants (not necessarily equal at all vertices), these eigenstates remain unaltered.
In particular, if the remaining network is open and has a continuous spectrum, the constructed eigenfunctions will be BICs.
This conclusion applies only to the subset of ring eigenstates pertaining to the appropriate symmetry class.
All other states of the subnet will be strongly mixed with the states of the remaining network and in the large network limit the overwhelming majority of states will be ergodically distributed over the whole network, as expected for typical systems with wave chaos \cite{Haake2018}.

The mechanism described differs strongly from the generation of BICs via global symmetries.
Indeed, there is no symmetry protection here as the environment of the subnet is asymmetric.
Rather, the local symmetry guarantees the existence of states with zero amplitude at the coupling points and destructive interference between all outgoing waves.
Instead of any unguided fine-tuning of parameters there is a clear design principle allowing this situation to be realized.
So in a sense the local symmetry combines features of accidental and symmetry-based BICs into a new mechanism.
This mechanism is independent of specific properties of the network such as the precise boundary conditions at the vertices or their valency (= the number of connected edges).
Most importantly, it does not require any symmetry of the network as a whole.

Note that the term {\it local} symmetry does not imply a subnet with a small total bond-length; rather, it refers to the fact that it involves only a subset of connected edges.
In particular, the BIC might be supported by a subnet that connects distant parts of the total network either by involving a few long edges or many short ones.

While pure BIC states are completely decoupled and do not contribute to transport across the network, any small perturbation of the subnet will create quasi-BICs which act as channels across the network and thus strongly affect its transport properties.

\begin{figure*}
  \centering \includegraphics[width=\linewidth]{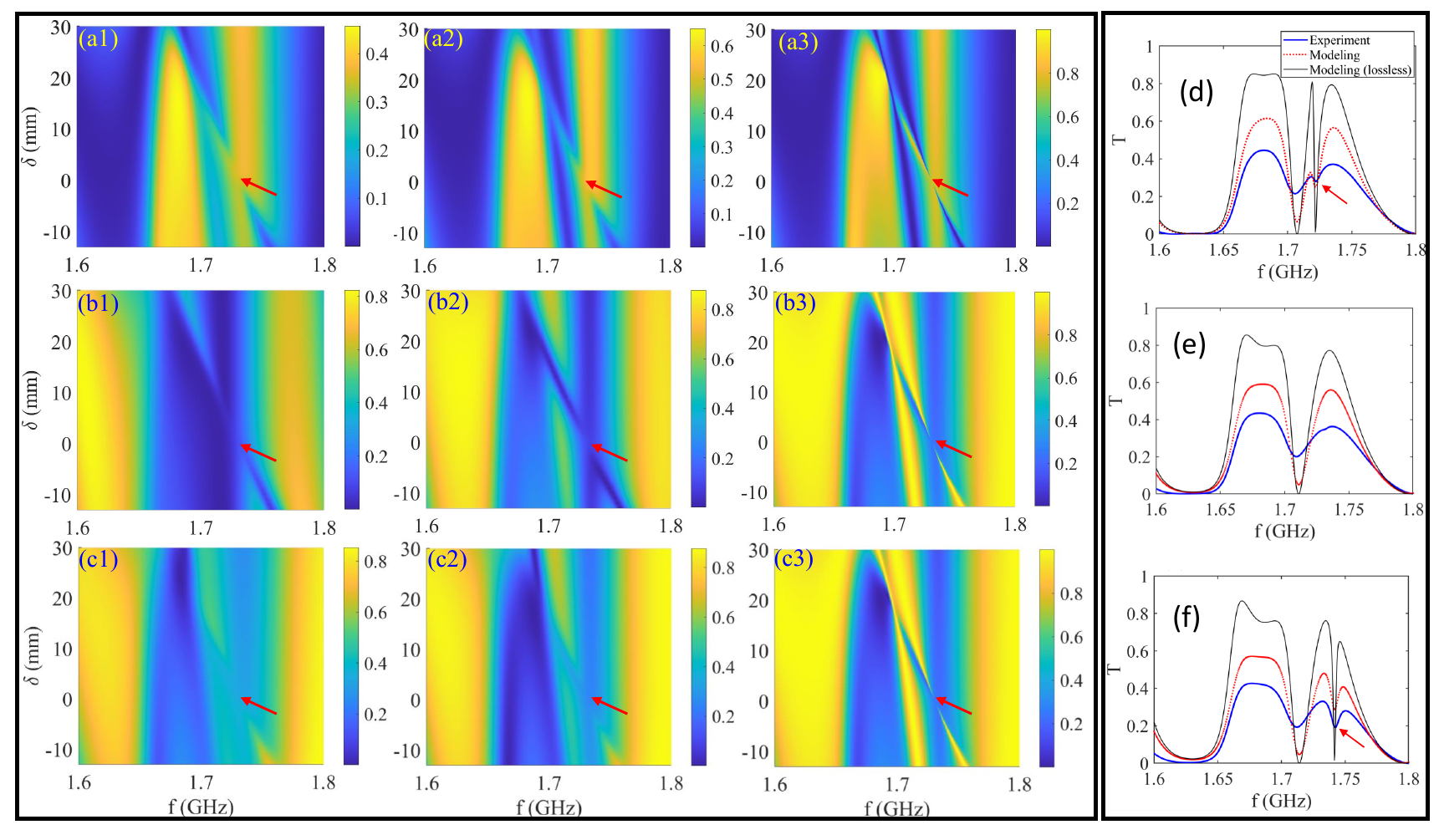}
  \caption{ \label{fig:TandRat1.7GHz}
		{\bf The transmittance and reflectance spectrum versus the phase-shifter length variation $\delta$ and frequency $f$.}
(a1) The experimental measurements of the transmittance through the graph (input from lead $1$, transmittance is the same as that from lead $2$ due to reciprocity) versus the input frequency and phase-shifter length variation $\delta$.
(a2) and (a3) the same as in (a1) but for the theoretical modeling with adjusted loss and without loss, respectively.
(b1)-(b3) The reflectance from lead 1 versus the input frequency and phase-shifter variation $\delta$ for experimental measurement, theoretical modeling with loss, and theoretical modeling without loss,  respectively.
(c1)-(c3) The same plot as in (b1)-(b3) for reflectance from lead 2.
(d)-(f) The cross-section plot of transmittance $T$ for $\delta=6$, $0$, and $-6$\,mm respectively.
The blue solid, red dotted, and black solid lines are for the experimental measurement, theoretical modeling with fitted loss, and theoretical modeling without loss, respectively.
}
\end{figure*}

\section{Transport in Complex Networks and BIC formation}\label{sec:Transport}\vspace{-0.25cm}

\paragraph*{Scattering on Complex Networks ---}
We study the scattering on a complex network of $n=1,\cdots,N$ vertices, where two vertices $n,m$ may be connected by an edge $E=(n,m)$ with length $l_E$.
The position $x_E=x$ on edge $E$ is $x=0(l_E)$ on vertex $n(m)$.
The wave $\psi_E(x)$ on the edge $E$ satisfies the Helmholtz equation
\begin{equation}\label{Helm}
	\frac{d^2}{d x^2}\psi_E(x)+ k^2\psi_E(x)= 0\,,
\end{equation}
where $k=\omega n_r/c_0$ is the wave number, $\omega$ is the angular frequency, $c_0$ is the speed of light, and $n_r$ is the complex-valued relative index of refraction that includes the losses of the coaxial cables.
The solution to Eq.~(\ref{Helm}) is $\psi_{E}(x) = \phi_{n}\frac{\sin k(l_{E} - x)}{\sin kl_{E}} + \phi_{m} \frac{\sin kx}{\sin kl_{E}}$, where $\psi_E(0)=\phi_{n}$ and $\psi_E(l_E)=\phi_{m}$ are the values of the field at the vertices.
We turn the compact network into a scattering setup set-up by attaching transmission lines (TLs) $\alpha=1,\cdots, L$ to a subset of the vertices.
The field on the $\alpha$th TL takes the form $\psi_{\alpha}(x)={\cal I}_\alpha e^{-ikx}+{\cal O}_{\alpha}e^{+ikx}$ for $x\ge 0$, where $x=0$ is the position of the vertex.
The coefficients ${\cal I}_\alpha ({\cal O}_{\alpha})$ indicate the amplitude of the incoming (outgoing) wave on the TLs.
At each vertex $n$, continuity of the wave and current conservation are satisfied.
These conditions can be expressed in a compact form as \cite{kottos2000chaotic}
\begin{equation} \label{GMM}
	(M + iW^TW)\Phi = 2iW^T{\cal I}\,,
\end{equation}
where $\Phi = (\phi_1,\phi_2,\cdots, \phi_{N})^T$.
The $L$ dimensional vector ${\cal I}$ contains the amplitudes ${\cal I}_\alpha$ of the incident field, while $W$ is an $L\times N$ matrix describing the connection between the TLs and the vertices.
A matrix element $W_{\alpha,n}$ is 1, if the $\alpha$th TL is attached to vertex $n$ and 0 otherwise.
The $N \times N$ matrix $M$,
\begin{eqnarray} \label{Mmatrix}
	M_{nm} =\left\{
	\begin{array}{ll}
		-\sum_{l\neq n}{\cal A}_{nl}\cot kl_{nl}, & \textrm{if } n = m\\
		{\cal A}_{nm}\csc kl_{nm}, &\textrm{if } n \neq m
	\end{array}\right.
\end{eqnarray}
incorporates information about the metric (length of edges) and the connectivity of the network, where ${\cal A}$ is the adjacency matrix having elements ${\cal A}_{nm}=1$ whenever two vertices $m,n$ are connected and ${\cal A}_{nm}=0$ otherwise.

The scattering field $\Phi$ on the compact part of the graph can be evaluated by solving Eq.~(\ref{GMM}) for $\Phi$.
The same expression, together with the continuity condition at the vertices, where the TLs are attached, can be used for deriving the scattering matrix \cite{kottos2000chaotic}
\begin{equation}\label{Smatrix}
  S(k) = -\hat{I} + 2iW\frac{1}{M(k) + iW^TW}W^T\,.
\end{equation}
For wave numbers which are integer multiples of $\pi/l_{nl}$ the terms $M_{nm}$ diverge; this can be rectified by appropriate manipulation of the divergent terms; see \cite{KS99}.

The poles of the scattering matrix in the complex  $k$-plane are related to resonances (i.e.\ purely outgoing solutions of the wave equation) and they are found from the condition $\det(M(k_p) + iW^TW) = 0$.
Of interest are also the zeros of the scattering matrix defined via the secular equation $\det S(k_z)=0$.
For lossless structures, causality implies  $k_z=k_p^*$.
The zeros $S(k)=0$ correspond to a special type of wave fronts with time-modulated amplitude, known as coherent virtual absorption, which are temporarily trapped inside the structure without any leakage \cite{Baranov+2017,Longhi+2018}.
When Ohmic losses are also included, one can find parameters of the structure for which the complex zeros cross the real axis.
In this case, there are stationary, perfectly impedance-matched input wave fronts that are completely absorbed by the (weakly) lossy elements in the structure which acts as an interferometric trap, known as coherent perfect absorber \cite{Chong+2010}.
The exceptional case where the poles are equal to the zeros of the $S$-matrix corresponds to BIC states which contain neither an incoming nor an outgoing radiation component and exist at a real frequency of passive structures.
Thus a BIC is invariant under time reversal and it does not affect the on-shell scattering matrix $S(k)$ since it is decoupled from the far-field radiation.
In the topological-defect picture in the complex $k$-plane a BIC implies that a charge $+1$ ($S$-matrix zero) annihilates with a charge  $-1$ (resonance) on the real axis.
We have confirmed this topological feature experimentally (see below).

\begin{figure*}
  \centering \includegraphics[width=\linewidth]{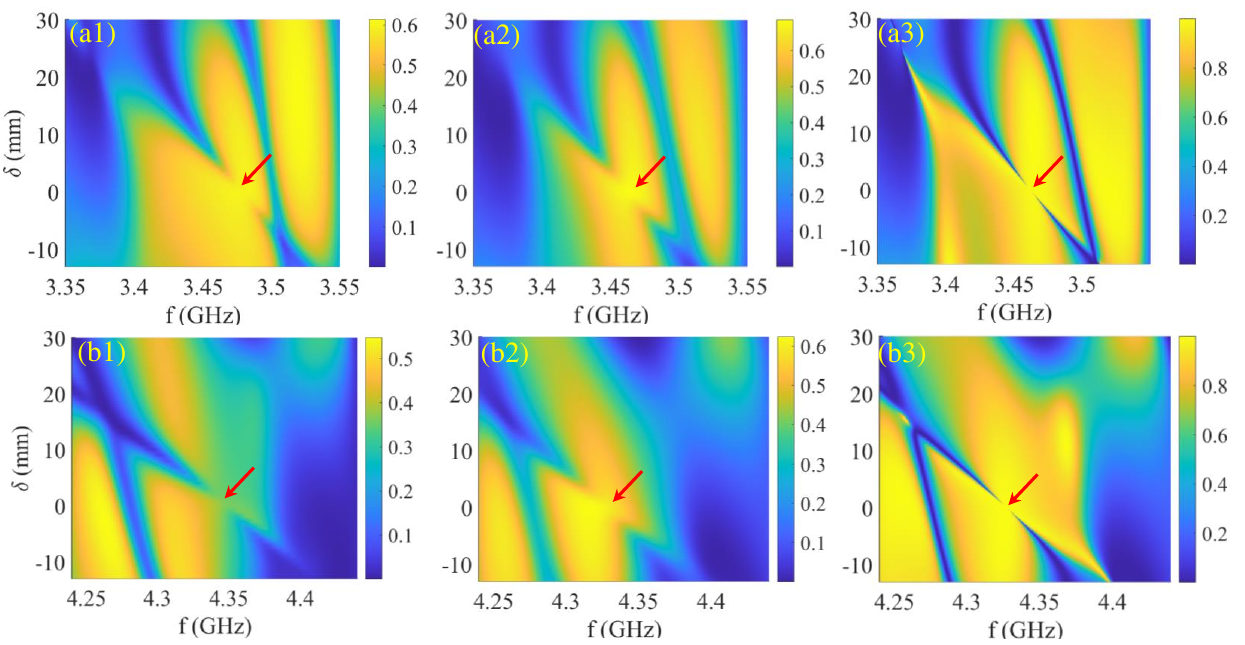}
  \caption{ \label{fig:Tfmultiple}
  	{\bf Transmittance spectra at frequencies that are multiples of the fundamental BIC frequency.}
	(a1)-(a3) The transmittance as a function of input frequency and bond variation $\delta$ for the experimental measurement (a1), the theoretical modeling with loss (a2) and without loss (a3) for the fourth BIC state at $4 \times 0.86547$\,GHz $\approx 3.46$\,GHz).
	(b1)-(b3) The same as in (a1)-(a3) but for the fifth BIC state at $4.33$\,GHz.
}
\end{figure*}

\paragraph*{BIC and quasi-BIC states --}
We now identify the conditions that are required for the realization of locally symmetric BIC states on a complex network.
To this end, we consider a subgraph consisting of  $C$ edges that form a closed loop within the network [for example, in Fig.~\ref{fig:ExpSetup}(a), $C=5$ for the violet pentagon].
We assume that they have equal lengths $\ell$ and construct a state $\psi_{c}(x)=\sqrt{(2/C\ell)}\sin(k_{M}x)$ which is restricted to this cycle.
This requires $k_{M}=M\pi/\ell$ and $k_{M}=2M\pi/\ell$ ($M=1,2\dots$) for even and odd $C$, respectively.
It is easy to verify that this state is normalized and that continuity and current conservation are satisfied at all vertices.
Thus, it is an eigenstate of the network.
This argument is completely independent of the topology of the rest of the network, the number of extra edges that are attached to the vertices of the cycle, and the number of attached TLs.
An example of a quasi-BIC state and its progression towards a BIC (as parameters of the network are changed) is shown in Fig.~\ref{fig:wave-dist} of Appendix~\ref{app:TheoBICModes}.
Further considerations reveal that it is possible to relax the assumption of equal edge lengths $\ell$ on the subgraph to rationally related lengths.

In an experiment, the BICs are manifested as long-lived quasi-bound (resonant) states that disappear and reappear in a characteristic manner when the edge lengths along the subgraph are changed by a small amount $\delta$.
For the unperturbed BIC state the wave function on an edge is a sine wave $\sin(kx)$ with nodes at two vertices.
When the edge length is changed, a linear approximation to the sine function results in an amplitude $\psi_{c}(0)\sim k\delta$ on the vertices.
It can be shown that such a perturbation gives rise to a Feshbach resonance with width $\gamma \sim|\delta|^{2}$ and a linear shift in the wave number or frequency, $\Delta f\sim \delta$.
The latter reveals itself as a sharp feature in the transmittance $T(f)$ and left (right) reflectance $R_1(f)$ ($R_2(f)$), where $R_1=R_2$ if there are no losses inside the network.
This feature is absent when $\delta=0$, and this is an indirect, but experimentally observable, signature of a BIC in scattering data.

Finally, we comment on the compatibility to the nonexistence theorem for BICs in compact structures mentioned above \cite{HZSJS16}.
The theorem has been derived under the requirement of smooth optical potentials where the wave function is analytic.
If such a function is identically zero in a limited region of space, this implies that it must vanish everywhere.
By contrast, in the model that we described above it is straightforward to construct eigenstates which have their support on a subset of the edges while being identically zero everywhere else.
This is possible as the boundary conditions require the wave function to be continuous but not analytic across a vertex (the derivative has a discontinuity for any vertex with degree larger than 2).
The validity of such boundary conditions in the modeling of photonic networks has been verified in many experiments \cite{H+04c,RAJMSKS16,Chen22,L+20,All+2014,CKA20}.
If no quasi-1D network model is to be used and rather the waveguides (microwave cables in our case) are modeled in 3D physical space, the theorem is still not applicable as in this case the boundaries of the waveguides correspond to infinitely high potential barriers that extend to infinity.

\section{Experimental Implementation}\vspace{-0.25cm}\label{sec:implementation}

We consider the tetrahedron network shown in Figs.~\ref{fig:ExpSetup}(b) and \ref{fig:ExpSetup}(c).
This network is relatively simple and does not have any geometrical symmetries.
At the same time, the dynamics is rich enough to show typical features of wave chaos \cite{Haake2018, kottos1999periodic,kottos2000chaotic,SK03}.
A microwave implementation is realized using coaxial cables (Huber+Suhner S~04272) connected by 6 T-junctions (vertices).
The electrical permittivity of the cables was found to be $\epsilon\approx 1.47(\pm0.07)+i0.0055(\pm0.0005)$ indicating the presence of uniform losses.
Two TLs have been attached at the vertices labeled $1$ and $2$; see Figs.~\ref{fig:ExpSetup}(b) and \ref{fig:ExpSetup}(c).

We choose the triangle consisting of vertices 4, 5, and 6 as the closed loop that supports BICs.
The electrical lengths $L_{45}=L_{46}=\ell=346.6$\,mm are fixed while the length $L_{56}$ of the third edge varies such that $L_{56}=\ell + \delta$ with $\delta\in [-14,30]$\,mm.
This length variation is done using a phase shifter whose effective length is controlled electronically.
The other lengths of the network are $L_{13}=L_{25}=18.2$\,mm, $L_{23} = 424.2$\,mm, $L_{16}=926.5$\,mm, and $L_{34}=831.4$\,mm.
See Appendix~\ref{app:ExpCharacterization} for details concerning the definition and determination of the electrical lengths.

Following the theoretical arguments of the previous section BICs will appear at $\delta=0$, for wave numbers $k_{M}=2M\pi/\ell$, $M \in \mathbb{N}^+$, corresponding to frequencies $f_M = M\frac{c_0}{\ell} = M \times 0.86547$\,GHz.
In the experimentally accessible frequency range $0.5-5$\,GHz we expected and observed a total of five BICs.
In the following, we report details for three of them, while the remaining two are  discussed in the Appendix~\ref{app:TheoBICModes}.

In Figs.~\ref{fig:TandRat1.7GHz}(a1), \ref{fig:TandRat1.7GHz}(b1), and \ref{fig:TandRat1.7GHz}(c1), we show the measured transmittance $T$ and reflectances $R_1$ and $R_2$ versus the input frequency and the tuning parameter $\delta$ in the proximity of the second BIC frequency ($2\times 0.86547$\,GHz $\approx 1.73$\,GHz) indicated by the arrow.
The second (third) column reports our calculations for the network of Fig.~\ref{fig:ExpSetup}(b) and \ref{fig:ExpSetup}(c) in the presence (absence) of Ohmic losses at the coaxial cables.
In the frequency range $f\in [1.6, 1.8]$\,GHz, a BIC is predicted at $f_2 \approx1.73$\,GHz.
In all cases we find a resonance moving linearly from around 1.7\,GHz at $\delta=15$\,mm to 1.76\,GHz at $\delta=-10$\,mm.
At $\delta=0$\,mm a BIC is formed and the resonance feature disappears from all three spectra ($T(f)$ and $R_{1,2}(f)$).
This is the expected indirect signature of the BIC: since the state is completely decoupled from the rest of the network and the TLs, any incident radiation cannot excite it and therefore there are no signatures of its existence in the scattering matrix elements.
Instead, for small $\delta\ne 0$, the transmittance $T(f\sim f_M)$ and reflectance $R(f\sim f_M)$ show a narrow resonance structure in their frequency dependence whose width is controlled by the parameter $\delta$.
All these features of the transmittance and reflectance spectra are present in our measurements and in both sets of calculations (with and without losses).
The presence of losses, however, smooths out some of the sharp characteristics of the (quasi-)BIC resonance (first and second columns) which are much more pronounced in the calculations shown in the third column where we have considered the same network without any losses of the coaxial cables.

To further investigate the variations of the resonance features in $T(f)$ as the tuning parameter changes around $\delta=0$ we plot in the right column of Fig.~\ref{fig:TandRat1.7GHz} the transmittance spectrum for three different $\delta$ values around the BIC value.
For $\delta=6$\,mm [Fig.~\ref{fig:TandRat1.7GHz}(d)] a narrow resonance dip (indicated by the red arrow) is evident in both, measurements (blue solid line) and calculations (red dotted line), where the Ohmic losses of the cables are taken into account.
This dip becomes very sharp in the case of a lossless network modeling (black solid line).
When $\delta=0$\,mm, [Fig.~\ref{fig:TandRat1.7GHz}(e)], the resonance dip disappears in all cases and reappears again when $\delta=-6$\,mm, [Fig.~\ref{fig:TandRat1.7GHz}(f)], after acquiring a small blueshift.

Beside Ohmic losses, imprecision in the lengths of the three cables can also influence the experimental manifestation of a BIC.
While a uniform change of the lengths will simply shift the observed feature along the frequency axis, in the case of independent variations $\varepsilon_{nm}$ the BIC condition cannot be restored with a single tunable parameter, i.e., a resonance will persist.
However, as outlined at the end of the previous section, the resonance width depends quadratically on the variation in the cable lengths, $\gamma\sim\delta^2$.
This contributes to the robustness of the experimental signature of a BIC with respect to noise in the cable lengths as errors smaller than the controlled variation $\varepsilon_{nm}\ll\delta$ will result in a much smaller resonance width which is not visible experimentally.
Note also that a variation of lengths outside the loop supporting the BIC is not critical as it affects the smooth background of the transmission spectrum but not the width and the location of the resonance peak.

We have also confirmed experimentally the appearance of a cascade of bound states at predetermined $k$-values which are multiples of a ``fundamental'' BIC frequency.
To demonstrate this feature we report as examples in Fig.~\ref{fig:Tfmultiple} the transmittance versus frequency and $\delta$ in the frequency regions where we expect the fourth and fifth BIC to occur, i.e., $f$=3.35-3.55\,GHz, and $f$=4.24-4.44\,GHz.
The same behavior as found for the BIC presented in Fig.~\ref{fig:TandRat1.7GHz} is observed here as well.
Namely, we observe the appearance and disappearance of a quasi-BIC mode as $\delta$ varies, signifying the formation of BICs at $\delta=0$.
While theoretically infinitely many BICs would occur for increasing frequency the number of experimentally observable states is limited by the frequency range of our network analyzer.
Besides the three BICs shown in Figs.~\ref{fig:TandRat1.7GHz} and \ref{fig:Tfmultiple}, there are two other BIC modes that are part of the BIC ladder in the frequency range of investigation.
The corresponding data are shown in Fig.~\ref{fig:BICladder} in the Appendix~\ref{app:ExpDemBICLadder}.

To conclude this section, let us discuss in some detail the difference between symmetry-induced and accidental BICs.
In both cases, destructive interference between outgoing waves is the basic mechanism for the decoupling of states.
Accidentally, a number of partial waves may cancel each other when a number of parameters are varied.
Lacking an underlying theoretical scheme, a prediction of such accidental cancelations is possible at most by numerical simulations, and they must take into account information from the entire network.
By contrast, symmetry-induced BICs (global or local symmetry) are guaranteed to occur for all eigenstates pertaining to a specific representation of the symmetry group.
So, unlike accidental BICs, they occur with a predictable pattern---an infinite periodic ladder of frequencies in our case (Fig.~\ref{fig:Tfmultiple}).
Moreover, it is sufficient to control the parameters of the locally symmetric substructure.
Hence, for a large network with a small symmetric subnet there is a vast reduction of free parameters, which need to be fine-tuned.
This means also that BICs based on a local symmetry are stable with respect to changes in the surrounding network.
We believe that in applications this can be a particularly important advantage over accidental BICs.

\section{Poles, zeros, and the topological structure of BICs}

\begin{figure*}
  \centering \includegraphics[width=0.8\linewidth]{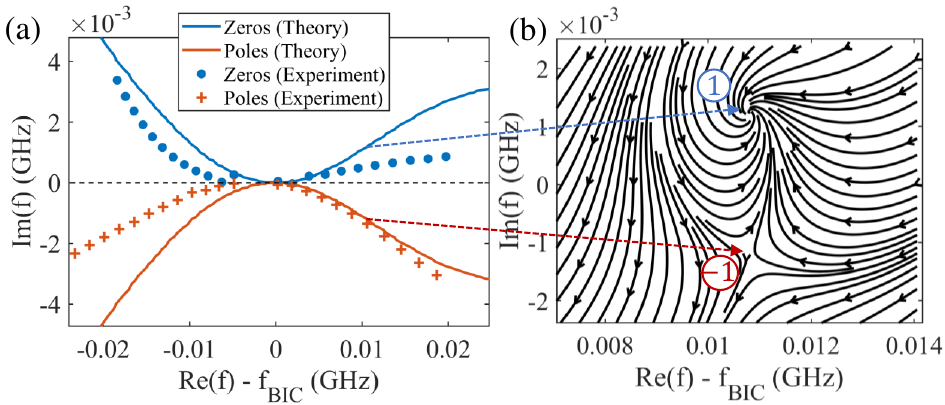}
  \caption{ \label{fig:poles-zeros}
	{\bf Formation of a BIC with local symmetry via the annihilation of a zero and a pole pair.} (a) Parametric evolution of poles (red) and zeros (blue) of the scattering matrix over the cable length $l_{56}$ of the tetrahedron network of Fig.~\ref{fig:ExpSetup}(b) and \ref{fig:ExpSetup}(c) for $\delta\in [-12, 12]$\,mm.
    The experimental poles (red crosses) and zeros (blue-filled circles) have been extracted from the measured $S$-matrix.
    The numerical results are indicated with lines of the respective color and have been extracted from the $S$-matrix given by Eq.~\ref{Smatrix}.
    All data have been shifted along the ${\cal I}m(k)$ axis by $0.00314$ due to the global loss induced by the Ohmic resistances in the coaxial cables.
    $f_{\rm{BIC}}  = 1.7309$\,GHz corresponding to the BIC in Fig.~\ref{fig:TandRat1.7GHz}.
    (b) Streamlines of the vector field $\vec{\Psi}=\nabla_{\vec{k}}arg\left(\det(S(\vec{k})\right)$ with $\vec{k}=(\Re(k),\Im(k))$, showing $+1$ and $-1$ topological charges at zeros and poles of the $S$-matrix corresponding to a $\delta=-6$\,mm.
    }
\end{figure*}

Finally, we analyze the formation of BICs from a topological perspective \cite{SHS20} by studying the parametric evolution of the poles and zeros in the complex frequency plane as $\delta$ is changing.
We point out that the increase/decrease of the electrical length of the cable introduces additional variations in the Ohmic losses which also contribute to the motion of the zeros and poles in the complex plane.
In Fig.~\ref{fig:poles-zeros}(a) the symbols represent the poles (red crosses) and zeros (blue-filled circles) of the $S$-matrix that have been extracted from the measured $S$-matrix using the harmonic inversion method \cite{kuh08b}.
The method was applied to the off-diagonal elements of the $S$-matrix for the poles and to the matrix $S^{-1}$ for the zeros.
The solid lines correspond to the results of the simulations.
The slight discrepancy between the experimental data and the numerical evaluation of the imaginary parts of zeros and poles is associated with an unavoidable uncertainty in the experimental procedure that was used for the extraction of the cable losses that have been used in our modeling.

Note the parabolic shape of the curve in the lower complex plane.
This confirms the quadratic dependence of the resonance width on the distortion $\gamma\sim\delta^2$ as the imaginary part of the pole corresponds to the resonance width while the shift in the real part is linear (see Sec.~\ref{sec:Transport}).

Our analysis indicates the coalescence of the poles and zeros at $\delta=0$ and at the BIC frequency $f_{\rm BIC}$ as discussed above (see Sec.~\ref{sec:Transport}).
Specifically, the formation of the BIC in the complex ${\vec k}=\left(\Re(k);\Im(k)\right)$ plane, is a consequence of the annihilation of two topological defects of the secular function $\det S({\vec k})$ \cite{SHS20}: a zero ${\vec k}_z$ and a pole ${\vec k}_p$ of the scattering matrix which collide as the perturbation parameter $\delta$ vanishes [see Fig.~\ref{fig:poles-zeros}(a)].
The zero (pole) is characterized by the topological charge $q$=+1 (-1), where
\begin{equation} \label{qcharge}
q\equiv\frac{1}{2\pi} \oint d{\vec k} \cdot \nabla_k\phi({\vec k})
\end{equation}
describes how many times the total phase of the scattering matrix $\phi(k)=\arg(\det S(k))$ winds by $2\pi$ along a counterclockwise simple closed path that encloses the topological defect [see Fig.~\ref{fig:poles-zeros}(b)].
A nonzero charge $q\neq 0$ indicates that a zero or a pole cannot suddenly disappear when the parameter $\delta$ slightly varies, although they can move in the complex ${\vec k}$-plane.
Only a collision of a $+1$ charge with a $-1$ charge can result in their mutual annihilation, which signifies that at this parameter value $\delta$ a resonance mode, i.e., a solution of the wave operator with outgoing boundary conditions, is also an eigenmode of the wave operator with incoming boundary conditions signifying a zero mode.

\section{Conclusion}\vspace{-0.25cm}

We have implemented a physical mechanism based on local symmetries that leads to the creation of BICs in compact photonic networks without any geometrical symmetry.
The resulting BICs are formed as a consequence of the collision of two topological defects (a pole and zero of the scattering matrix) which leads to their annihilation.
The proposed BICs differ from existing scenarios, where BICs are formed due to the presence of a global symmetry or where they are the "accidental" result of a parameter variation.
In particular, they do not require the degeneracy of two (or more) resonant modes and they are based on a precise rule that allows the construction and control of a ladder of BIC states at multiples of a fundamental frequency.

The proposed BICs are robust to uniform losses and perturbations that maintain the nodal structure of the BIC mode.
Furthermore, any perturbation outside the local structure does not affect their formation.
Instead, they are extremely sensitive to frequency detuning and variations of the electric length (either via index or physical length variations) of the subdomain that supports the BICs---a property that can be used for highly efficient sensing (where the sensing platform is the local-symmetry subdomain).
These variations can be also self-induced via nonlinear interactions occurring along the bonds of the subdomain and can be enabled by the injected light intensities which act as a tuning parameter for tunable channel dropping and light storage and release \cite{krasikov2018nonlinear,BPS13,BPS15}.

In the optical framework, we envision applications of our approach in a variety of light technologies that rely on enhanced nonlinear light-matter interaction effects.
For example, the high-Q characteristics of the local-symmetry-based BICs can have ramifications for the development of BIC-based lasers for on-chip integrated coherent light sources, second and third harmonic generation, four-wave mixing, sensing, etc.\ (for recent reviews see \cite{HZSJS16,JPSJ21,AK21}).

Our microwave proof-of-principle demonstration of local-symmetry-based BICs can be utilized for secure wired communications \cite{PDH23}. It can also be extended to optical photonic platforms.
Promising candidates are coupled fiber networks \cite{Lepri17} that have recently been used to demonstrate wave-chaotic properties of complex networks like the one used in this paper or photonic integrated microring resonators with engineered coupling between them
\cite{LHHJJCK21}.
We also envision the implementation of such local-symmetry-based BICs to other wave platforms ranging from acoustics \cite{RRTMSHA23} to elastodynamics \cite{BCT18}.

Acknowledgement - CZW, AD and TK acknowledge partial support from Grant No.~ECCS2148318 within the Resilient \& Intelligent NextG Systems (RINGS) program and from Simons Foundation for Collaboration in MPS Grant No.~733698.
UK and HS acknowledge the hospitality of Wesleyan University.

\appendix
\begin{figure*}
  \centering \includegraphics[width=\linewidth]{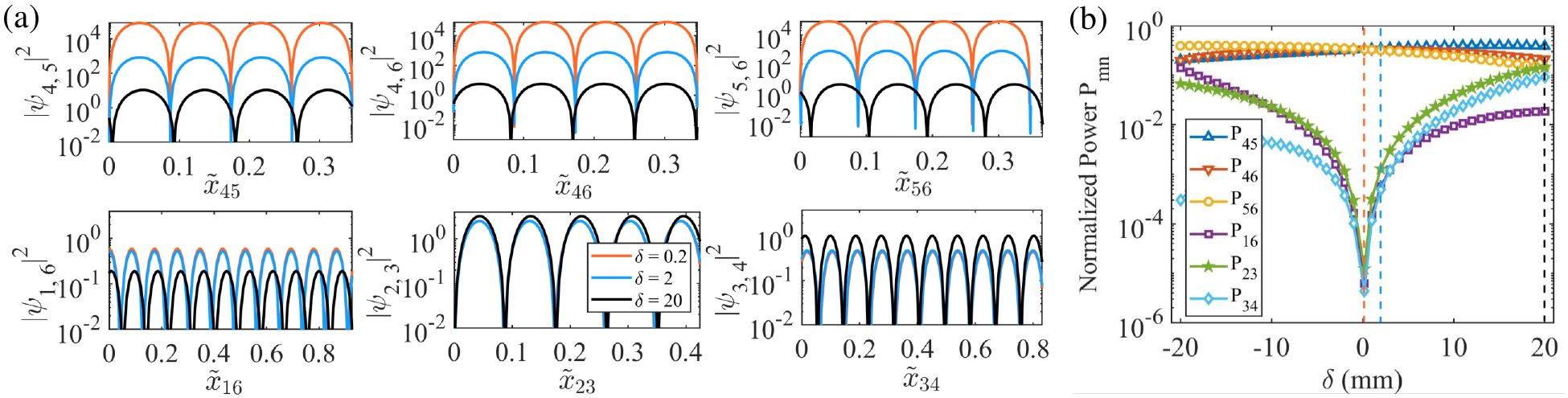}
  \caption{ \label{fig:wave-dist}
    {\bf Intensity distribution of BIC states.}
    (a) Intensity distribution of the scattering field at the resonance frequency, versus the electric distance ${\tilde x}=n x$ along the bonds that form the network of Fig.~\ref{fig:ExpSetup}(b) and \ref{fig:ExpSetup}(c).
    The local symmetry $C_{3v}$ that is responsible for the formation of a BIC is imposed on the subnet that consists of the vertices $n=4,5,6$, and occurs when the detuning parameter $\delta=0$.
    As $\delta$ increases the $C_{3v}$ is violated and the BIC is destroyed.
    Three values of the detuning parameter are used:
    $\delta=0.2$\,mm (red lines), $\delta=2$\,mm (blue lines), and $\delta=20$\,mm (black lines).
    We observe that for $\delta\rightarrow 0$ the field at the bonds that form the subnet increases dramatically indicating the formation of a BIC.
    (b) The normalized power at each bond, versus the detuning parameter $\delta$.
    }
\end{figure*}

\section{Theoretical evaluation of the spatial distribution of BIC modes}
\label{app:TheoBICModes}

To get a better understanding of the formation of BICs and their spatial structure inside a network, we calculate the scattering fields.
For presentation purposes, we consider the simple case of a lossless network and assume the scenario where the incident wave is injected into the structure from lead~1.
Use of Eq.~(\ref{GMM}) allows us to obtain the wave amplitudes $\Phi=(\phi_1,\phi_2,\cdots,\phi_N)^T$ on each of the $n=1,2,\cdots,N$ vertices.
The scattering field on a bond $L_{nm}$ can be expressed in terms of $\{\phi_n\}$ as $\psi_{E}(x) = \phi_{n}\frac{\sin k(l_{E} - x)}{\sin kl_{E}} + \phi_{m} \frac{\sin kx}{\sin kl_{E}}$, where $\psi_E(0)=\phi_{n}$ and $\psi_E(l_E)=\phi_{m}$ are the values of the field at the vertices.
We introduce a length-detuning $\delta$ ($L_{56}= \ell + \delta$) on the bond connecting vertex 5 to 6.

In Fig.~\ref{fig:wave-dist}(a) we report the scattering field intensities at each bond for the network of Fig.~\ref{fig:ExpSetup}(b) and \ref{fig:ExpSetup}(c).
Three different detunings $\delta=0.2$, $2$, and $20$ have been used in order to demonstrate the progression towards the second BIC (see Fig.~\ref{fig:TandRat1.7GHz} of the main text).
For these calculations, we have used the resonance frequencies, associated with the resonance frequency of transmittance which are $1.730$, $1.727$, $1.706$\,GHz respectively.
For presentation purposes, we have chosen to report the field versus the electric distance ${\tilde x}=n_r\cdot x$ which is the same for all three bonds of the subdomain where a $C_{3v}$ hidden symmetry is imposed.
(we do not present the plotting of the $L_{13}$ and $L_{25}$ bonds since they are very short).
We see that the intensity is dramatically enhanced on the bonds $L_{45}$, $L_{46}$, and $L_{56}$ while the field amplitudes $\phi_n,\phi_m$ on the $n,m$ vertices (edge points) decreases as the detuning $\delta$ gets smaller.
This trend signifies the formation of a BIC mode occurring at $\delta=0$.
In this case, the field acquires the form $\psi_{c}(x)=\sqrt{(2/3\ell)}\sin(k_{M}x)$ at all bonds of the subdomain and zero anywhere else (see the discussion in Sec.~\ref{sec:Transport}.

To further analyze the BIC modes, we calculate the normalized power $P_{nm}=\int_0^{L_{nm}}|\psi_E(x)|^2/\sum_E\int_0^{L_{nm}}|\psi_E(x)|^2$ on each bond $E=(n,m)$, and plot it as a function of detuning $\delta$; see Fig.~\ref{fig:wave-dist}(b).
We see that as $\delta$ decreases towards zero, the normalized power increases on the bonds of the subdomain that supports the BIC state and decreases to zero on all other bonds.

\section{Experimental Characterization of the Complex Network}
\label{app:ExpCharacterization}

To determine the parameters entering the numerical simulation, we have measured the different cable lengths as $l_{16}=726$\,mm, $l_{34}=644$\,mm, $l_{23} = 310$\,mm, and $l_{45}=l_{46}=245.7$\,mm.
Bear in mind that the real length is also includes lengths from the T-junctions which are 18\,mm.

The electrical length of the cables and the phase shifter are defined as the product of the measured geometrical length and the real part of the refractive index, $L=\Re(n)\,l$.
We extract the complex refractive index $n$ of the cables and phase shifter based on the transmission measurements.
Comparing the wave propagation through the coaxial cable to the Helmholtz equation for a 1D free and uniform medium, we can get the transmission coefficient $t$ as
\begin{equation}
t = \frac{\psi(l)}{A} = e^{i\frac{\omega}{c}\Re(n)l}e^{-\frac{\omega}{c}\Im(n)l}\,,
\end{equation}
where we have the refractive index as $n=\sqrt{\epsilon_0}$, the angular frequency $\omega=2\pi f$, and the cable length $l$.
First, we can get the real part of $t$ as
\begin{equation}
  \Re(t) = \cos\left[\frac{\omega}{c}\Re(n)l\right]\,e^{-\frac{\omega}{c}\Im(n)l}\,,
\end{equation}
where we can see that the cosine term is the oscillation part, which could give us the $\Re(n)$ based on our measured $t$ in a frequency range and measured cable length $l$.
Second, the imaginary part of refractive index $n$ can be obtained based on the transmittance $T=|t|^2$ which is of the form
\begin{equation}
  T = |t|^2=e^{-\frac{2\omega}{c}\Im(n)l}\,,
\end{equation}
where we can get $\Im(n)$ by fitting the curve of transmittance versus frequency for a certain cable length $l$.
To avoid the influence of the SMA connector effect on the cable length, in our measurement, we use the ratio of transmissions $t_1/t_2$ for two cables with different lengths $l_1$, $l_2$.
In this way, we only need the cable length difference so that we can get rid of the error from the SMA connector and obtain more accurate values for $\Re(n)$ and $\Im(n)$.
By fitting the transmission curves, we get $\Re(n) = 1.212$ and $\Im(n)=0.0022$ for coaxial cables.
For the phase shifter, we obtain the refractive index as $\Re(n)=1.004$ and $\Im(n)=0.002$.
Note, that $\Im(n)$ is frequency dependent,
which we ignored thus taking only an average $n$ for the whole investigated frequency range into account.

The phase shifter is adjusted such that its electrical length is equivalent to the electrical lengths of the other two cables in the symmetric triangle, $L_{45}=L_{46}$, when we have $\delta=0$\,mm corresponding to a fixed real length of the phase shifter.
An increase of $\delta$ corresponds to increasing the electrical length of the phase shifter.

The electrical lengths which are based on geometric measurements cannot account for additional phase shifts at the connections and inside the T-junctions and for the uncertainty in the extracted permittivities of the cables.
Therefore, for our numerical modeling of the network, we refine the parameters usingan optimization procedure.
Starting from the measured lengths of the cables and the T-junctions as well as the complex refractive index, we allow for a fluctuation range for each variable.
Then we use the surrogate optimization method in MATLAB to find the parameter set that gives the best fit for the transmittance and reflectance curves over the whole accessible frequency range $0.5-6.5$\,GHz.
The fitted geometrical lengths are $\hat l_{16} = 764.4$\,mm, $\hat l_{34} = 686$\,mm, $\hat l_{23} = 350$\,mm, $\hat l_{45}=\hat l_{46} = 286$\,mm, $\hat l_{13} = \hat l_{25} = 15$\,mm, and the refractive index for the cables is $\hat n_c = 1.212+0.0022i$.
The electrical lengths used in the simulations and quoted in Sec.~\ref{sec:implementation} are the product of the optimized geometrical lengths and the optimized value for the real part of the refractive index, $L=\Re(\hat n_c)\,\hat l$.
The optimized refractive index for the phase shifter is $\hat n_p = 1.0 + 0.002i$ and the electrical length is $\hat{l}_{56}=286{\rm\,mm} \cdot \Re(\hat n_c) + \delta \cdot \hat n_p$, with $\delta$ the detuning parameter.

\section{Experimental demonstration of the existence of a BIC ladder}
\label{app:ExpDemBICLadder}

\begin{figure*}
  \centering \includegraphics[width=0.8\linewidth]{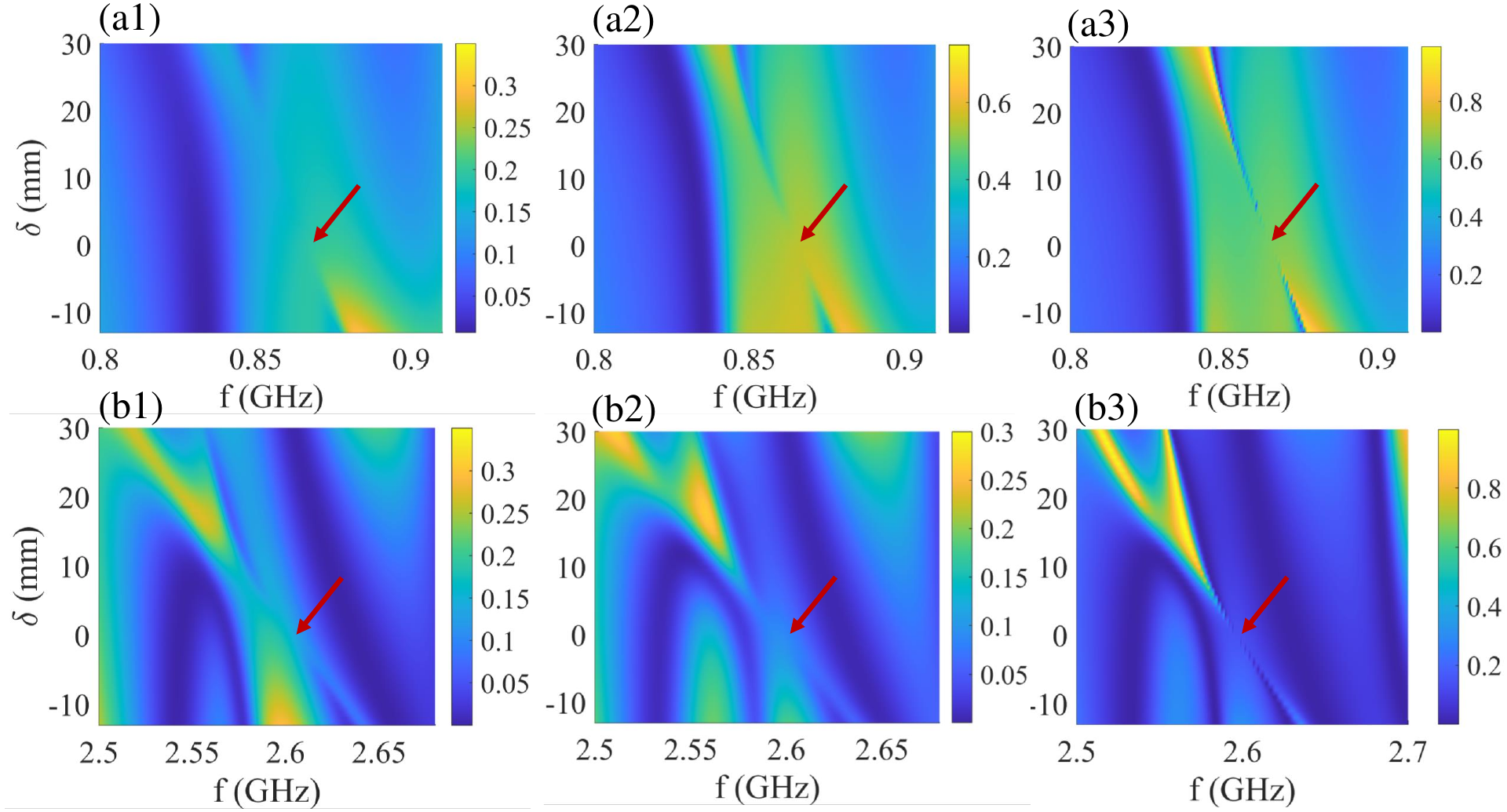}
  \caption{ \label{fig:BICladder}
    {\bf Formation of a BIC ladder.} (a1)-(a3) The transmittance as a function of input frequency and bond variation $\delta$ for the experimental measurement (a1), the theoretical modeling with loss
    (a2) and without loss (a3) for the first (fundamental) BIC state at $f=0.86547$\,GHz.
    (b1)-(b3) The same as in (a1)-(a3) but for the third BIC state at $f=3\times 0.86547 $\,GHz.
    }
\end{figure*}

We have also investigated the formation of the fundamental $(f=0.86547)$\,GHz and the third $(f=3\times 0.86547)$\, GHz BIC states associated with the network of Figs.~\ref{fig:ExpSetup}(b) and \ref{fig:ExpSetup}(c).
These two BIC modes, together with the three BICs presented in the main text, constitute the first five modes in the BIC ladder, which were experimentally accessible.

Figures~\ref{fig:BICladder}(a1) and \ref{fig:BICladder}(b1) show the measured transmittance $T$ versus frequency $f$ and bond detuning $\delta$ in the frequency range where we expect the formation of the fundamental and third BIC states, respectively.
The experimental results compare nicely with the numerically evaluated transmittances, in the case of lossy networks (Figs.~\ref{fig:BICladder}(a2) and \ref{fig:BICladder}(b2) where the complex index of refraction that is used in our modeling has been extracted using the method described in Appendix~\ref{app:ExpCharacterization}.
We have completed our analysis by evaluating the transmittances for the same network system in the absence of losses; see Figs.~\ref{fig:BICladder}(a3) and \ref{fig:BICladder}(b3).
In complete analogy with the results shown in Figs.~\ref{fig:TandRat1.7GHz},\ref{fig:Tfmultiple} of the main text, we observe again the appearance and disappearance of a quasi-BIC mode as $\delta$ varies, signifying the formation of a BIC at $\delta =0$.
In the right column (lossless network), one can further recognize the sharp characteristics of BIC modes, which are otherwise masked by the losses (first and second columns of Fig.~\ref{fig:BICladder}).
Note, that in all five cases the BIC behavior is the same but the background reflections and transmissions is different as they do not respect the $C_{3v}$ symmetry leading to different interference patterns of the BIC-related resonances which gives rise to Fano-like lineshapes of the BIC resonance.

%

\end{document}